\documentclass[aps,prl,preprint,superscriptaddress]{revtex4-1}

\usepackage{graphicx}
\usepackage{upgreek}
\usepackage{siunitx}

\usepackage{units}
\usepackage{textcomp}
\usepackage[yyyymmdd,hhmmss]{datetime}

\usepackage{amsmath}
\usepackage{amssymb}

\usepackage{xcolor}

\usepackage[final]{hyperref}
\definecolor{link}{rgb}{0,0,0.9}
\definecolor{cite}{rgb}{0,0,0.9}
\definecolor{url}{rgb}{0,0,0.9}
\hypersetup
{
    colorlinks=true,
    linkcolor=link,
    citecolor=cite,
    urlcolor=url,
}

\usepackage{soul}
\definecolor{disc}{rgb}{0.9,1,0.9}

\definecolor{new}{rgb}{1,1,0.9}

\newcommand{\VTPI}{V_\text{TPI}}

\newcommand{\iu}{{i\mkern1mu}}
\newcommand{\abs}[1]{\left|#1\right|}
\newcommand{\fig}[1]{Figure #1}
\newcommand{\eq}[1]{equation (\ref{#1})}

\begin{document}

\title{Two-photon interference in the telecom C-band after frequency conversion of photons from remote quantum emitters}

\author{Jonas H. Weber}
\thanks{These authors contributed equally}
\affiliation{Institut f\"ur Halbleiteroptik und Funktionelle Grenzfl\"achen, Center for Integrated Quantum Science and Technology (IQ{$^{ST}$}) and SCoPE, University of Stuttgart, Allmandring 3, 70569 Stuttgart, Germany}
\author{Benjamin Kambs}
\thanks{These authors contributed equally}\affiliation{Fachrichtung Physik, Universit\"at des Saarlandes, Campus E 2.6, 66123 Saarbr\"ucken, Germany} 
\author{Jan Kettler}\affiliation{Institut f\"ur Halbleiteroptik und Funktionelle Grenzfl\"achen, Center for Integrated Quantum Science and Technology (IQ{$^{ST}$}) and SCoPE, University of Stuttgart, Allmandring 3, 70569 Stuttgart, Germany}
\author{Simon Kern}\affiliation{Institut f\"ur Halbleiteroptik und Funktionelle Grenzfl\"achen, Center for Integrated Quantum Science and Technology (IQ{$^{ST}$}) and SCoPE, University of Stuttgart, Allmandring 3, 70569 Stuttgart, Germany}
\author{Julian Maisch}\affiliation{Institut f\"ur Halbleiteroptik und Funktionelle Grenzfl\"achen, Center for Integrated Quantum Science and Technology (IQ{$^{ST}$}) and SCoPE, University of Stuttgart, Allmandring 3, 70569 Stuttgart, Germany}
\author{H\"useyin Vural}\affiliation{Institut f\"ur Halbleiteroptik und Funktionelle Grenzfl\"achen, Center for Integrated Quantum Science and Technology (IQ{$^{ST}$}) and SCoPE, University of Stuttgart, Allmandring 3, 70569 Stuttgart, Germany}
\author{Michael Jetter}\affiliation{Institut f\"ur Halbleiteroptik und Funktionelle Grenzfl\"achen, Center for Integrated Quantum Science and Technology (IQ{$^{ST}$}) and SCoPE, University of Stuttgart, Allmandring 3, 70569 Stuttgart, Germany}
\author{Simone L. Portalupi}\email{s.portalupi@ihfg.uni-stuttgart.de}
\affiliation{Institut f\"ur Halbleiteroptik und Funktionelle Grenzfl\"achen, Center for Integrated Quantum Science and Technology (IQ{$^{ST}$}) and SCoPE, University of Stuttgart, Allmandring 3, 70569 Stuttgart, Germany}
\author{Christoph Becher}
\affiliation{Fachrichtung Physik, Universit\"at des Saarlandes, Campus E 2.6, 66123 Saarbr\"ucken, Germany} 
\author{Peter Michler}
\email{p.michler@ihfg.uni-stuttgart.de}
\affiliation{Institut f\"ur Halbleiteroptik und Funktionelle Grenzfl\"achen, Center for Integrated Quantum Science and Technology (IQ{$^{ST}$}) and SCoPE, University of Stuttgart, Allmandring 3, 70569 Stuttgart, Germany}
\homepage[]{www.ihfg.physik.uni-stuttgart.de}

\begin{abstract}
\textbf{
Efficient fiber-based long-distance quantum communication via quantum repeaters relies on deterministic single-photon sources at telecom wavelengths, with the potential to exploit the existing world-wide infrastructures. For upscaling the experimental complexity in quantum networking, two-photon interference (TPI) of remote non-classical emitters in the low-loss telecom bands is of utmost importance. With respect to TPI of distinct emitters, several experiments have been conducted, e.g., using trapped atoms \cite{Beugnon2006}, ions \cite{Maunz2007}, NV-centers \cite{Bernien2012, Sipahigil2012}, SiV-centers \cite{Sipahigil2014}, organic molecules \cite{Lettow2010} and semiconductor quantum dots (QDs) \cite{Patel2010,Flagg2010,He2013b,Gold2014,Giesz2015,Thoma2017,Reindl2017,Zopf2017}; however, the spectral range was far from the highly desirable telecom C-band. Here, we report on TPI at $1550 \,$nm between down-converted single photons from remote QDs \cite{Michler2017Book}, demonstrating quantum frequency conversion \cite{Zaske2012,Ates2012,Kambs2016} as precise and stable mechanism to erase the frequency difference between independent emitters. On resonance, a TPI-visibility of \SI{29+-3}{\percent} has been observed, being only limited by spectral diffusion processes of the individual QDs \cite{Robinson2000,Kuhlmann2015}. Up to 2-km of additional fiber channel has been introduced in both or individual signal paths with no influence on TPI-visibility, proving negligible photon wave-packet distortion. The present experiment is conducted within a local fiber network covering several rooms between two floors of the building. Our studies pave the way to establish long-distance entanglement distribution between remote solid-state emitters including interfaces with various quantum hybrid systems \cite{DeGreve2012,Maring2017,Bock2017,Maring2018}.
}
\end{abstract}
\maketitle

\begin{figure*}
	\centering
	\includegraphics{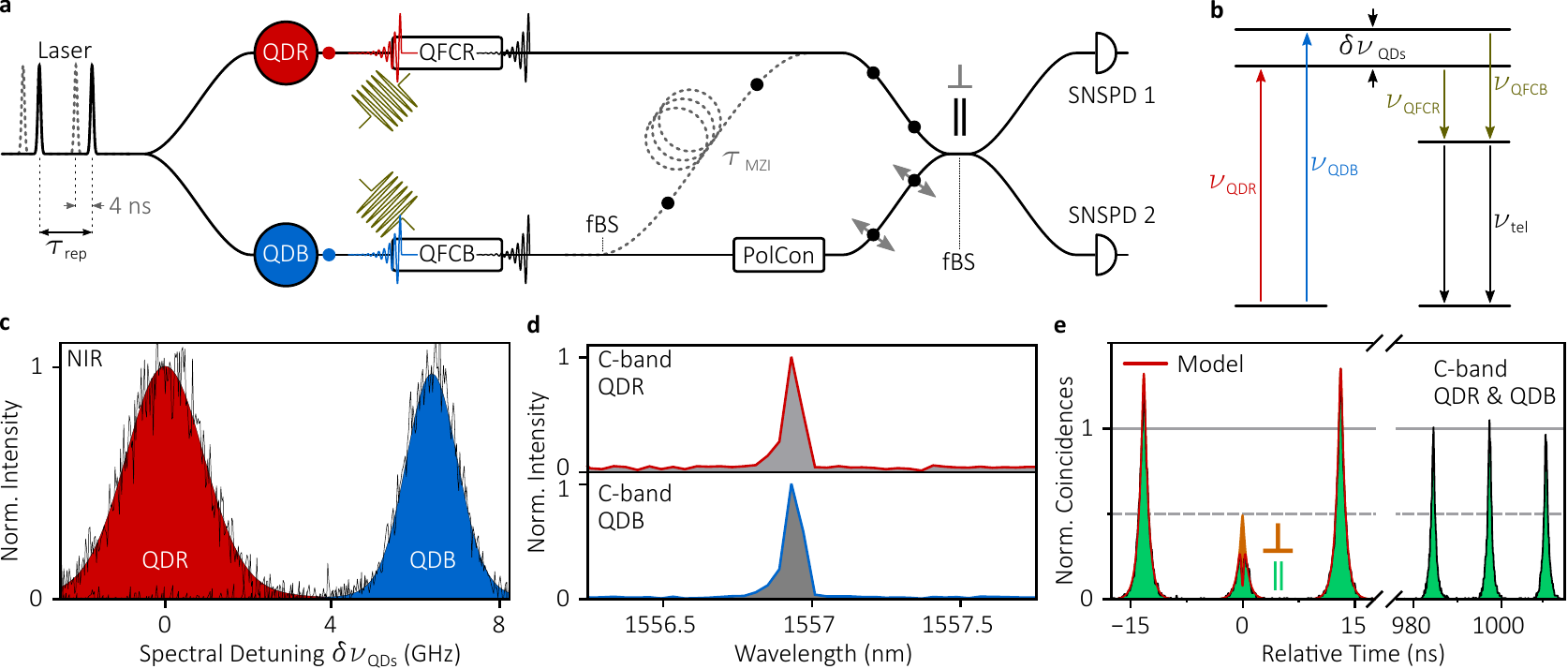}
	\caption{(a) Sketch of experimental setup utilized for TPI measurements of consecutively emitted photons (dotted line configuration for QDB) as well as TPI with remote QDs situated in distinct cryostats (solid line configuration). In both configurations, the emission is transferred to the telecom C-band via two distinct frequency converters named QFCR and QFCB. A polarization control (PolCon) is used to realize distinguishable ($\perp$) and indistinguishable ($\parallel$) photon overlap on a fiber-based beamsplitter (fBS). Detected events are time tagged for the following correlation analysis. (b) Schematic energy diagram for the QFDC: the converters are used to precisely control the spectral matching. (c) High-resolution PL (resolution $\approx \SI{100}{\mega\hertz}$) spectra of the QD emission in the NIR allowing for reliable setting of the two converters. (d) PL spectra in the telecom C-band after QFDC (standard resolution $\approx \SI{10}{\giga\hertz}$). (e) Remote TPI measurement of QDR and QDB with parallel and orthogonal polarization setting. Central peaks and temporally distant peaks (setting the Poissonian level) are displayed.}
	\label{fig:fig1}
\end{figure*}

Perspective quantum repeater scenarios include quantum interference of single photons -- preferably at telecom wavelengths -- from remote quantum nodes \cite{Sangouard2007}. In this respect, two-photon interference represents the basic quantum operation to establish entanglement between different nodes and therefore several efforts have been made to improve the maximally achieved TPI visibility. Among non-classical light emitters, QDs reach near-ideal values using photons from the same emitter \cite{Michler2017Book}. However, when interfering photons stemming from remote emitters, the achieved visibilities are well below unity, mainly limited by spectral diffusion \cite{Loredo2016, Wang2016}. Despite this challenge, TPI from remote emitters represents a viable approach for all applications where active demultiplexing cannot be employed (i.e. mid- and long-distance quantum networks) \cite{Wang2017}. In the telecom C-band, current state-of-the-art showed TPI with one down-converted quantum emitter and a laser \cite{Yu2015,Felle2015}. As a clear step forward, we here demonstrate TPI with on-demand generated photons of two distinct remote quantum dots. By means of two independent quantum frequency converters (QFCs) we transfer NIR-photons to the telecom C-band and at the same time exploit quantum frequency down-conversion (QFDC) as highly stable and precise tuning mechanism to overcome the spectral offset of the utilized remote QD-pair (here denoted as QDR:QDB). Figure~\ref{fig:fig1}a depicts the experimental setup, where the same pulsed laser is used to excite the two emitters situated in separate cryostats. Single photons are generated by resonantly addressing the charged exciton transitions via coherent $\pi$-pulse excitation and then forwarded to the QFCs. As the relative detuning of the pump lasers is set to compensate the frequency mismatch of the original near infrared photons, the retrieved telecom photons are brought into resonance (compare Figure~\ref{fig:fig1}b). After QFDC of both photon streams in individual frequency converters, the telecom photons are sent via a \SI{60}{\meter}-fiber link to another floor of the building. The investigated system represents a model of a realistic scenario where optical fibers are crossing several different rooms instead of a controlled lab environment. At the end of the fiber network, the non-classical photons are brought to interference, feeding a fiber-based beamsplitter (fBS).

The two QDs utilized in the experiment show a TPI visibility of $V _\text{QDR,NIR}^\text{\SI{4}{\nano\second}}=\SI{72+-4}{\percent}$ and $V _\text{QDB,NIR}^\text{\SI{4}{\nano\second}}=\SI{58+-4}{\percent}$, respectively. The frequency detuning $\delta\nu_\text{QDs}$ between the s-shells (compare high-resolution PL (hPL), in Figure~\ref{fig:fig1}c) can be compensated via temperature tuning, allowing to frequency match the two emitters. It is worth noting that when performing remote TPI, here with QDR:QDB, the interfering photons are spectrally completely uncorrelated \cite{Legero2003}, i.e., the maximum obtainable TPI visibility is determined by both homogeneous (inferred from decay time measurement, see Suppl. Info.) and inhomogeneous broadening. For this reason, the measured remote TPI visibility in the NIR regime gives a state-of-the-art value of $V_\text{exp,NIR}^\text{remote}  = \SI{26+-3}{\percent}$ (see Ref.~\cite{Weber2017}). In the present study, QFDC is used to bridge the gap between NIR and telecom regime, then working with photons at $\approx\SI{1550}{nm}$ (Fig.~\ref{fig:fig1}d). In this regard, the sources' individual TPI visibilities after QFDC result in $V_\text{QDR}^\text{\SI{4}{\nano\second}}=\SI{64+-21}{\percent}$ and $V_\text{QDB}^\text{\SI{4}{\nano\second}}=\SI{60+-3}{\percent}$ (see Suppl. Info.), proving that QFDC preserves both photon coherence and temporal shape \cite{Zaske2012}. By utilizing the converter to fine-tune the photon energy and erase the initial frequency mismatch, TPI of telecom photons from remote emitters is conducted. With this proof-of-principle experiment, a maximum visibility of $V_\text{exp}^\text{remote}  = \SI{29+-3}{\percent}$ is observed (Figure~\ref{fig:fig1}e), in agreement with theoretical modeling of the data based on Ref.~\cite{Kambs2018} (see Methods), giving $V_\text{sim}^\text{remote} = \SI{27+-1}{\percent}$ as the highest achievable value. The very good correspondence with NIR-experiments, further shows that the two independently operated QFCs do not induce any measurable dephasing.

\begin{figure}
	\centering
	\includegraphics{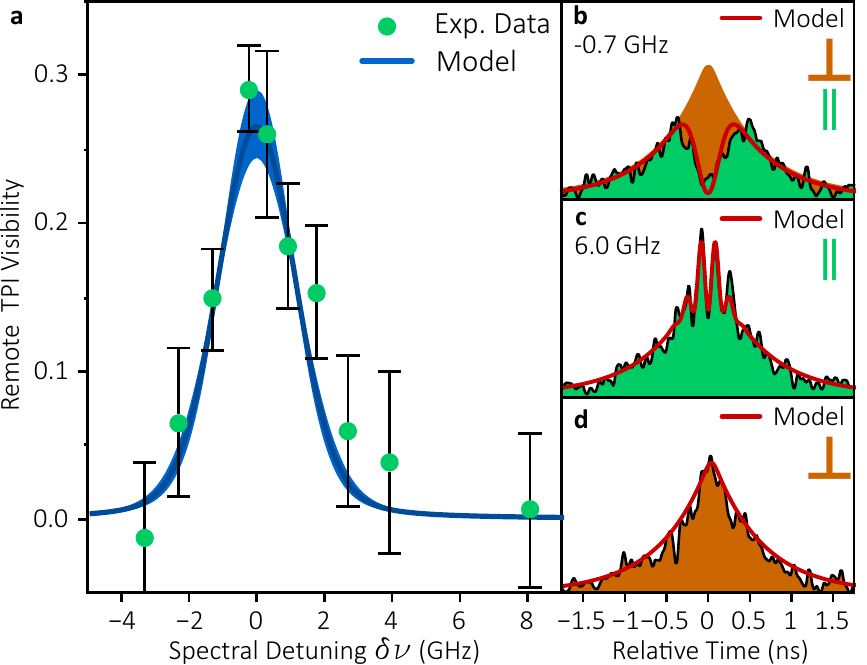}
	\caption{Remote TPI measurement with QDR and QDB. (a) TPI visibility in dependence of spectral detuning, obtained by varying the pump laser frequency of QFCB. Experimental data obtained via normalization on Poissonian level. Simulation is based on measurements of decay time and homogeneous broadening (see Methods). (b-c) High-resolution correlation measurements for parallel polarization and different spectral detuning ($\delta\nu = \SI{-0.7+-0.1}{\giga\hertz}$ and $\delta\nu = \SI{6+-0.1}{\giga\hertz}$ respectively). (d) Correlation measurement for orthogonally polarized photons.}
	\label{fig:fig2}
\end{figure}

In the following, the spectral offset between the two remote sources is precisely controlled using two independent QFCs: the accurate readout of the two pump laser frequencies allows for a stable and reliable variation of the spectral matching. To demonstrate the reliability of the tuning mechanism, TPI experiments for different converted wavelength detunings are performed. Figure~\ref{fig:fig2}a shows the visibility values in a frequency range of around \SI{12}{\giga\hertz}, however having available an overall tuning of more than \SI{2}{\tera\hertz}. The resulting tuning series of the remote TPI measurement proves the convenience and stability (\SI{20}{\mega\hertz}, i.e., orders of magnitude smaller than the natural linewidth, see Suppl. Info.) of the combined technologies with very good agreement between data and theoretical expectation. As can be seen in Figure~\ref{fig:fig2}b-d, the applied model fits very well with the obtained data. Panel~b shows the characteristic interference dip for spectral detuning of $\delta\nu = \SI{-0.7+-0.1}{\giga\hertz}$. Furthermore, based on the high time-resolution in these particular measurements, panel~c shows clear signatures of quantum beating in the center peak for spectral detuning of $\delta\nu = \SI{6+-0.1}{\giga\hertz}$, corresponding to a beating period of $170\,$ps. This effect was so far only shown for TPI with two remote atoms \cite{Legero2003} and organic molecules \cite{Lettow2010}. In panel~d no quantum interference can be observed as the polarization is set to be orthogonal, representing the distinguishable case. 

\begin{figure}
	\centering
	\includegraphics{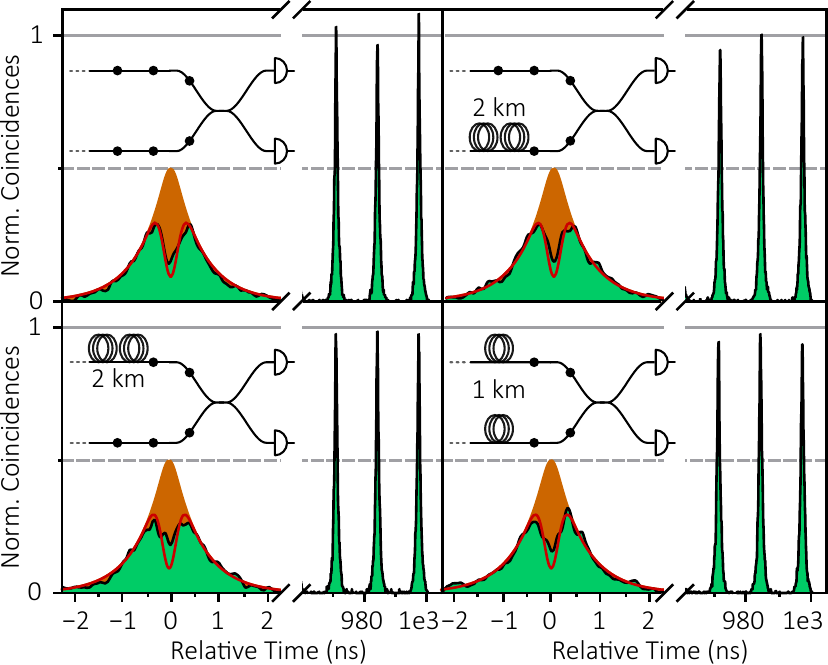}
	\caption{Demonstration of building-to-building network via additional fiber path length before TPI is carried out. Symmetric (0:0\,km and 1:1\,km) and asymmetric (2:0\,km and 0:2\,km) fiber configuration in the two emitter arms QDR:QDB. Coincidences are normalized to Poissonian level. Within the measurement error, no reduction of visibility is observed in all configurations.}
	\label{fig:fig3}
\end{figure}

Although the telecom C-band corresponds to the region of maximal transmission into a glass fiber, the spectral dispersion acquired during propagation of the single photons \cite{Lenhard2016} could still affect the TPI visibility over long distances. To further investigate the effect of photon wave packet distortion in a realistic building-to-building fiber-network scenario, remote TPI with additional fiber path length of up to \SI{2}{\kilo\meter} was carried out. Figure~\ref{fig:fig3} shows the resulting remote TPI measurements with symmetric (0:\SI{0}{\kilo\meter} and 1:\SI{1}{\kilo\meter}) as well as asymmetric (0:\SI{2}{\kilo\meter} and 2:\SI{0}{\kilo\meter}) fiber configuration between the photon streams of QDR:QDB. In the symmetric fiber configuration no reduction of TPI visibility is expected as the wave packets of both photon streams are equally affected by the material dispersion. The expectation is proven by the experimental results giving respectively $V^\text{0:\SI{0}{\kilo\meter}}=\SI{26+-2}{\percent}$ and $V^\text{1:\SI{1}{\kilo\meter}}=\SI{24+-2}{\percent}$, having a frequency detuning between the converted emitters of $\delta\nu = \SI{0.6+-0.1}{\giga\hertz}$. In case of an asymmetric configuration, dispersion may have a stronger effect on the wave packets traveling through the longer fiber. As a consequence, the mismatch in wave packet overlap would be increased leading to degradation of photon indistinguishability. However, in the experiment no reduction in TPI contrast can be observed for asymmetric fiber configuration, respectively ($V^\text{0:\SI{2}{\kilo\meter}}=\SI{22+-2}{\percent}$ and $V^\text{2:\SI{0}{\kilo\meter}}=\SI{23+-2}{\percent}$). 

\begin{figure}
	\centering
	\includegraphics{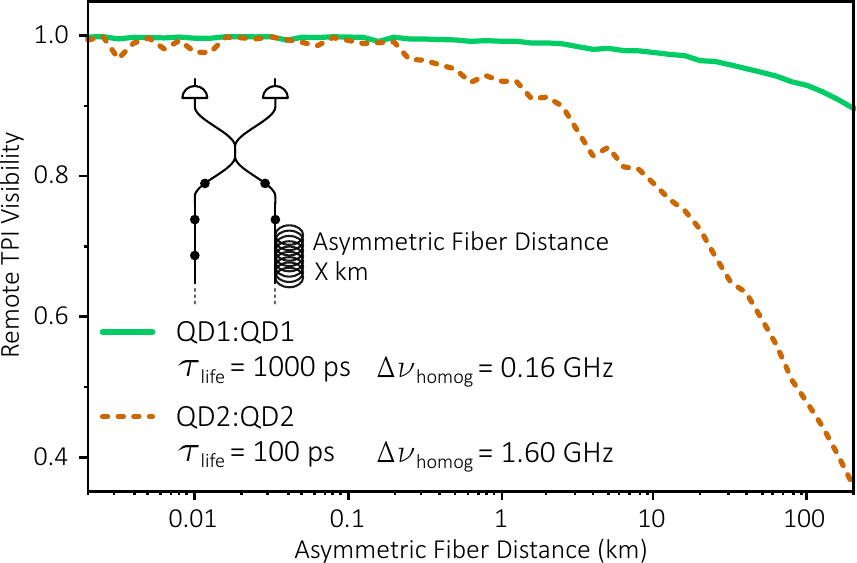}
	\caption{Photon wave-packet overlap for two remote QD scenarios after transmission through asymmetric fiber distance, respectively (see sketch in inset). In both scenarios two equal remote sources (QD1:QD1 and QD2:QD2) are simulated. The emitted photons are set to be Fourier-transform limited, however, having different lifetimes, i.e., bandwidth ($\tau_\text{life,QD1} = \SI{1000}{\pico\second}$, i.e., $\Delta\nu_\text{homog} = 0.16\,$GHz and $\tau_\text{life,QD2} = \SI{100}{\pico\second}$, i.e., $\Delta\nu_\text{homog} = 1.60\,$GHz). In each scenario, the photon stream of one of the two QDs is subject to a varying fiber length, to model a more realistic scheme where the sources are at different distance with respect to nodes.}
	 \label{fig:fig4}
\end{figure}

\noindent These results prove that for building-to-building fiber networks (up to \SI{2}{\kilo\meter}), the wave packet dispersion does not degrade the interference process, when operating in the telecom C-band. Nevertheless, in case of town-to-town fiber networks, the effects of dispersion may become non-negligible. It is further expected that via the time-bandwidth product of the Fourier-transform limited single photons \cite{Kuhlmann2015}, wave packet distortion should be stronger for shorter transition lifetimes corresponding to a larger spectral bandwidth. In order to quantify the visibility degradation in long-distance fiber networks, the simulations presented in Ref.~\cite{Vural2018} were applied. Figure~\ref{fig:fig4} shows the simulated remote TPI visibility of two ideal QD-pairs for asymmetric fiber configuration (0:X\,km), as this represents the limiting case scenario. Both QD-pairs have equal photon properties, respectively, i.e., they perfectly match in frequency and are set to have Fourier-transform limited integrated emission. To investigate the effect of different spectral bandwidths, the two pairs are set to have different lifetimes. Consequently, values which are typical for QDs embedded in planar structures \cite{He2013} and micro-cavities \cite{Somaschi2016} are chosen (QD1:QD1 with $\tau_\text{life} = \SI{1000}{\pico\second}$, i.e., $\Delta\nu_\text{homog} = 0.16\,$GHz and QD2:QD2 with $\tau_\text{life} = \SI{100}{\pico\second}$, i.e., $\Delta\nu_\text{homog} = 1.60\,$GHz). It is worth noting that while for QD-pair~1 the dispersion has a very limited effect even for long fiber length difference, for QD-pair~2 the remote TPI visibility drops significantly. For a realistic fiber path length difference of \SIrange{10}{100}{\kilo\meter} short transition lifetimes will then result in a \SIrange{20}{60}{\percent} drop of visibility comparing to \SIrange{2}{10}{\percent} variation for the case of QD-pair~1. Here, it becomes clear that when sources have to be used in fiber-based long-distance applications, care must be taken in adapting the emitter lifetime accordingly to the network design.

In conclusion, we demonstrated for the first time remote two photon interference in the telecom C-band using two distinct quantum emitters: exploiting quantum frequency conversion the NIR photons were transferred to $1550\,$nm  wavelength without compromising the photon quality, in terms of photon purity and indistinguishability. The presented experimental configuration further demonstrated that the utilized converters represent a very precise, stable and reliable mechanism to tune remote sources in resonance. The observed TPI contrast over spectral detuning shows very good agreement with the theoretical model and it compares well with state-of-the-art results for non-converted sources. The measurements additionally prove that the wave packet dispersion appearing while propagating into a glass fiber does not affect the visibility in a few-kilometer building-to-building network. The applied simulations further show that in case of asymmetric fiber path lengths, the remote TPI visibility is strongly decreased when working with short transition lifetimes; operating long-distance quantum networks, the emitter bandwidth has to be chosen carefully, depending on the fiber-length and the emitter counterpart. The described study represents a first key step forward in the implementation of realistic fiber-based quantum networks, clearly underlining the boundary conditions required for an effective implementation of such a highly desired quantum technology.

\section{Methods}
\subsection*{Experimental configuration}
The transition line of both QDs (see Ref.~\cite{Portalupi2016} for further info. on QD growth) is resonantly addressed at a repetition rate $1/\tau_\text{rep}= \SI{76.2}{\mega\hertz}$ with a pulse width of $\SI{3}{\pico\second}$. A confocal polarization suppression setup is utilized for top-excitation with a laser suppression of $\approx 10^7$. Both signals are send through a monochromator which is based on a transmission grating showing a spectral resolution of approximately \SI{15}{\giga\hertz}. To realize TPI, a fBS as well as a polarization control (PolCon) is utilized, giving \SI{99+-1}{\percent} of first-order interference visibility. Time-synchronization is carried out via variation of fiber and free-space delays in the excitation and emission paths, enabling both TPI with photons from remote and individual QDs. High-efficient ($\eta_\text{SNSPD} \simeq \SI{35}{\percent}$) and high-resolution ($\Delta\tau_\text{SNSPD} \simeq \SI{100}{\pico\second}$) superconducting nanowire single-photon detectors (SNSPDs) are utilized to capture coincidence events obtained during TPI. Data in Figure \ref{fig:fig2}b-d were recorded using detector with higher temporal resolution ($\Delta\tau_\text{SNSPD} \simeq \SI{40}{\pico\second}$ and $\eta_\text{SNSPD} \simeq \SI{40}{\percent}$).

At the heart of both conversion setups is an actively temperature stabilized, magnesium oxide doped, periodically poled lithium niobate (MgO:PPLN) waveguide chip ({\itshape NTT electronics}) with allover 18 waveguides. All waveguides have a rectangular cross section of \unit[(11x10)]{\textmu m$^2$}  and a length of \unit[40]{mm}. In order to minimize coupling losses, the end facets have an anti-reflective (AR) coating transparent for all participating light-fields. Each conversion setup is equipped with a single-frequency tunable Cr$^{2+}$:ZnS/Se Laser ({\itshape IPG Photonics}) as pump light source. The converter pump laser exhibit wavelength at around \mbox{$\lambda_\text{QFC}$ = \unit[2157]{nm}}.  The converters reach a maximum external conversion efficiencies of \unit[34.7]{\%} and \unit[31.4]{\%} for QDR and QDB, respectively. Both pump laser frequencies fluctuate with an rms level of \mbox{3$\sigma$ = \unit[78]{MHz}}. 
The fluctuation of the pump laser detuning $\delta\nu_\text{pump}$ is as few as \mbox{3$\sigma$ = \unit[20]{MHz}} (compare Suppl. Info.), which equals to the resolution limit of the used wavemeter. 

The tuning curve in Figure \ref{fig:fig2}a was measured with a coincidence rate of \SIrange{6000}{7000}{\hertz} and averaged over \SIrange{70}{80}{\minute}. Data in Figure \ref{fig:fig3} was measured with a coincidence rate of \SIrange{8000}{9000}{\hertz} and averaged over \SIrange{70}{180}{\minute}. During all measurements around \unit[55]{events/s} can be attributed to detector dark counts, another \unit[500]{events/s} result from conversion noise. The environment of the QDs is stabilized via additional CW above-band excitation contributing around \SI{10}{\percent} to the actual resonance fluorescence.

\subsection*{Theoretical modeling}
Remote TPI interference as well as the visibility dependence over the emitter detuning is modeled as described (for further details see Suppl. Info. and Ref.~\cite{Kambs2018}). In good approximation the relaxation dynamics can be described by the spontaneous emission of single photons from an ideal two-level system. Accordingly, the photon wave-functions $\zeta_{1,2}\left(t\right)$ can be expressed as a mono-exponential decay with transition life time $\tau_i$ and frequency $\nu_i$. The derivation starts from a well-established formalism describing the HOM experimental situation with the photon fields $\zeta_i\left(t\right)$ at the two inputs of a BS, respectively \cite{Legero2003}. Therein, the probability $P\left(t_0,\tau\right)$ with which both input photons leave the BS through distinct output ports and become detected at times $t_0$ and $t_0 + \tau$ is given by
\begin{equation}
P\left( t_0, \tau \right) = \tfrac{1}{4}\left|\zeta_1\left(t_0+\tau\right)\zeta_2\left(t_0\right)-\zeta_2\left(t_0+\tau\right)\zeta_1\left(t_0\right)\right|^2.
\label{pjoint}
\end{equation}
Using the wave-functions $\zeta_i\left(t\right)$, this leads to
\begin{align}
	g^{\left(2\right)}\left(\tau\right) &= \int_{-\infty}^{+\infty}{P \left( t_0,\,\tau \right)\,\text{d}t_0} \nonumber \\
	&= \frac{1}{4\left(\tau_1+\tau_2\right)} \times\left(e^{-\frac{\abs{\tau}}{\tau_1}} + e^{-\frac{\abs{\tau}}{\tau_2}} - 2 \cdot e^{-\frac{\abs{\tau}}{2T}}\cos{\left(2\pi\Delta\nu\tau\right)}\right),
	\label{g2define}
\end{align}
where the carrier frequency displacement is described by $\Delta\nu= \nu_1 - \nu_2$ and $1/T = 1/\tau_1 + 1/\tau_2$. Spectral diffusion processes of both QDs are included as $\Sigma^2 = \sigma_1^2+\sigma_2^2$, where $\sigma_i = \text{FWHM}_i/(2\sqrt{2\log2})$. The detuning of both center emission frequencies reads $\delta\nu = \nu_{c,1} - \nu_{c,2}$.  Accordingly, the measured cross-correlation $\mathcal{G}^{\left(2\right)}\left(\tau\right)$ in the long-time limit is obtained by a weighted average taking account for Gaussian frequency distributions leading to
\begin{align}
	\mathcal{G}^{\left(2\right)}\left(\tau\right) \propto\left(e^{-\frac{\abs{\tau}}{\tau_1}} + e^{-\frac{\abs{\tau}}{\tau_2}} - 2 \cdot e^{-\frac{\abs{\tau}}{2T}}e^{-2\pi^2\Sigma^2\tau^2}\cos{\left(2\pi\delta\nu\tau\right)}\right).
	\label{eq:VTPI_analytical}	
\end{align}
This is the final result used to describe the central peaks around $\tau= 0$ for all remote TPI correlation measurements shown in the present work.\par
Then, the TPI visibility of two remote emitters is defined by $V = 1 - 2\,P$, where $P$ is the overall probability of both photons going separate ways after meeting at the BS. Accordingly, $P$ can be calculated by integrating Equation (\ref{eq:VTPI_analytical}) with respect to the relative time $\tau$. The visibility evaluates to

	\begin{align}\label{eq:VTPI}
	V_\text{TPI} &= \frac{1}{2\sqrt{2\pi}\,\Sigma (\tau_{1} +
			\tau_{2})}\left[\exp{\left(-\frac{\left(i/(2T) + 2\pi\,\delta\nu\right)^2}{8\pi^2\,\Sigma^2}\right)}\right.\nonumber\\
		&\cdot\left.\text{erfc}\left(\frac{1/(2T) - i\,2\pi\,\delta\nu}{2\pi\sqrt{2}\,\Sigma}\right) + \text{c.c.}\right].
	\end{align}
Thus, the visibility of a remote HOM experiment is determined by the joint spectral properties of both emitters. Equation (\ref{eq:VTPI}) is used in the present work to predict the experimentally achieved visibilities as function of the pump laser detuning for the remote HOM case.

\begin{acknowledgments}
\section{Acknowledgments}
The authors thank T. Herzog for the installation of the telecom fiber links. Furthermore, we thank M. Bock for helpful discussions and advice during preparation of the QFC setups. This work was financially supported by the DFG via the projects MI 500/26-1 and BE 2306/6-1 as well as by the German Federal Ministry of Science and Education (Bundesministerium f\"ur Bildung und Forschung (BMBF)) within the project Q.com (Contract No. 16KIS0115 and 16KIS0127).
\section{Author contributions}
J.H.W, B.K, S.K. and S.L.P. performed the experiment with the support of J.K.. B.K. built the frequency converters. M.J. provided the samples. J.H.W. and B.K analyzed the data. B.K., H.V., J.M. set up the theoretical model. J.M. and H.V. conducted the numeric simulations. J.H.W., S.L.P and B.K. wrote the manuscript with support of P.M. and input from all authors. P.M. and C.B. coordinated the project. All authors actively took part in all scientific discussion.  
\section{Additional information} 
Correspondence and requests for materials should be addressed to S.L.P. and P.M..
\section{Competing financial interests} 
The authors declare no competing financial interests.
\end{acknowledgments}

%

\section{Supplementary information to ``Two-photon interference in the telecom C-band after frequency conversion of photons from remote quantum emitters"}
\section{Quantum Frequency Converter}

\begin{figure}
	\centering
	\includegraphics{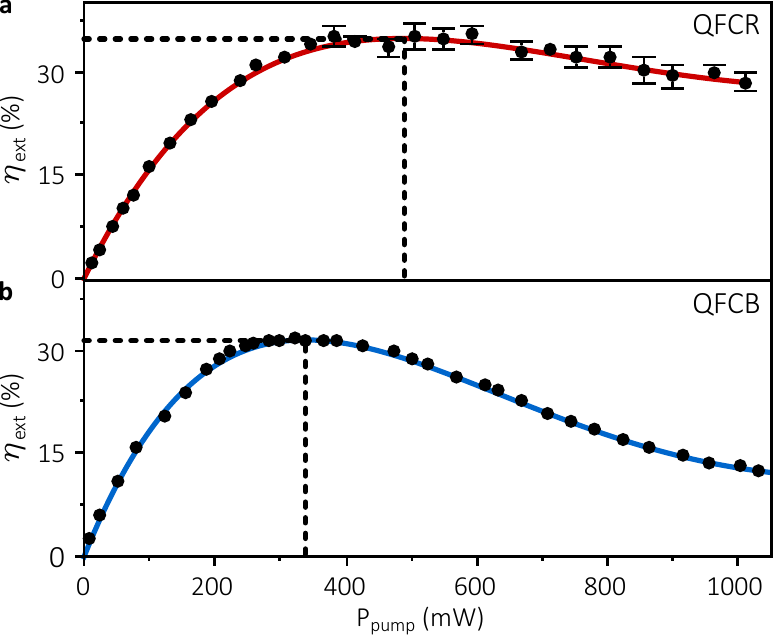}
	\caption{Converter efficiency versus pump laser power for the two independent QFCs of QDR (a) and QDB (b). The maximum conversion efficiency $\eta_\text{max}$ is marked as dotted line being  $\eta_\text{max}^\text{QFCR} = \SI{34.7}{\percent}$ and $\eta_\text{max}^\text{QFCB} = \SI{31.4}{\percent}$.}
	\label{fig:figS1}
\end{figure}

The single photons emitted by both quantum dots are independently fed into two identical, but distinct frequency converters. At the heart of both conversion setups is an actively temperature stabilized, magnesium oxide doped, periodically poled lithium niobate (MgO:PPLN) waveguide chip ({\itshape NTT electronics}) with allover 18 waveguides. All waveguides have a rectangular cross section of \unit[(11x10)]{\textmu m$^2$}  and a length of \unit[40]{mm}. In order to minimize coupling losses, the end facets have an anti-reflective (AR) coating transparent for all participating light-fields. The chip comes with 9 different poling-periods ranging from \unit[24.300]{\textmu m} to \unit[24.500]{\textmu m}. The periodic poling provides quasi-phase matching for a difference frequency generation (DFG) process transducing the input photons at \mbox{$\lambda_\text{QDR}$ = \unit[904.442]{nm}} and \mbox{$\lambda_\text{QDB}$ = \unit[904.420]{nm}} to the telecom C-band at \mbox{$\lambda_\text{tel}$ = \unit[1557.28]{nm}}. To achieve high conversion efficiencies, the process is stimulated by the presence of a pump light field, whose wavelength $\lambda_\text{p}$ fulfills the energy conservation relation of the DFG process \mbox{\unitfrac{1}{$\lambda_\text{p}$} = \unitfrac{1}{$\lambda_\text{QD}$} - \unitfrac{1}{$\lambda_\text{tel}$}}. This corresponds to \mbox{$\lambda_\text{p}$ = \unit[2157.46]{nm}} and \unit[2157.32]{nm} for QDR and QDB, respectively. Each conversion setup is equipped with a single-frequency tunable Cr$^{2+}$:ZnS/Se Laser ({\itshape IPG Photonics}) as pump light source. For power and polarization control, the pump beam passes a half-wave plate and a Glan-Taylor Calcite Polarizer. Both input light fields are spatially overlapped on a dichroic mirror and coupled to the waveguide via an aspherical zinc selenide lens with a focal length of \unit[11]{mm}. Subsequent to the conversion, the telecom photons are separated from the pump light by dichroic mirrors, coupled into a single-mode fiber and forwarded to analysis or further experiments. Due to anti-Stokes Raman scattering and a number of non-phase-matched nonlinear conversion processes acting on the pump light, a significant amount of background photons around the target wavelength are created whilst conversion. To minimize this unwanted contribution, the telecom photons are passed from a \unit[1550-20]{nm} bandpass filter as well as a system of a fiber circulator and a fiber Bragg grating, which acts as an additional \unit[121]{GHz} broad bandpass filter. At pump light powers of \unit[488]{mW} and \unit[338]{mW} the converters reach their maximal external conversion efficiencies of \unit[34.7]{\%} and \unit[31.4]{\%} for QDR and QDB as shown in Figure \ref{fig:figS1}. The external conversion efficiency was measured between converter input and output of the FBG filter stage and is defined as the fraction of usable converted photons over input photons.

\section*{Pump laser stability}
\begin{figure}
	\centering
	\includegraphics{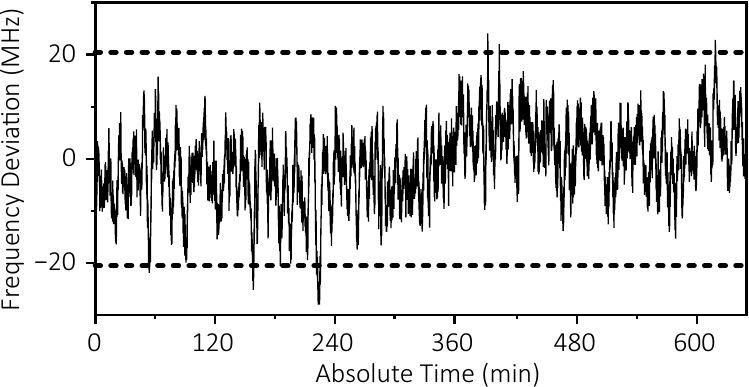}
	\caption{Relative pump laser stability $\delta\nu_\text{pump}$ -- around its mean value -- over time, showing a $3 \sigma$ deviation of $\pm\SI{20.3}{MHz}$.}
	\label{fig:figS2}
\end{figure}

The tuning precision of the telecom photons is of utmost importance. As described before, the wavelength of the output photons is set by the pump laser wavelength. As described before, the wavelength of the output photons is set by the pump laser wavelength, which can be controlled in a range from 2000\,nm to 2400\,nm with MHz precision. In order to monitor the output frequencies $f_\text{B}$ and $f_\text{R}$ of the pump lasers for conversion of QDB and QDR, the residual pump light is frequency doubled by means of a temperature stabilized MgO:PPLN bulk crystal subsequent to the conversion and then forwarded to a fiber based MEMS-switch. Here, both pump beams are combined before entering a wavemeter ({\itshape High Finesse, WS6-200}). To validate the desired detuning \mbox{$\delta\nu_\text{pump} = f_\text{R}-f_\text{B}$}, both output frequencies $f_\text{B}$ and $f_\text{R}$ were continuously monitored throughout all experiments. Additionally, the stability of the pump laser detuning $\delta\nu_\text{pump}$ was tested in a long-time  measurement \mbox{(approx. \unit[11]{h})} of $f_\text{B}$ and $f_\text{R}$. Both frequencies fluctuate with a rms level of \mbox{3$\sigma$ = \unit[78]{MHz}}. However, both lasers are operated at the same lab and therefore exposed to the same environmental conditions. Accordingly, their output wavelengths do not drift independently. As a result, the relative pump laser detuning $\delta\nu_\text{pump}$ fluctuates with a rms level as low as \mbox{3$\sigma$ = \unit[20]{MHz}} around its mean value $\langle  \delta\nu_\text{pump} \rangle$, which equals to the resolution limit of the used wavemeter (see Figure \ref{fig:figS2}). As this is around 2 orders of magnitude below the FWHM of the recorded tuning curve, it can be assumed that the conversion does not add any relevant frequency jitter to the telecom photons.

\section*{Theory}
In the present work, two-photon coalescence of single-photons emitted by two remote QDs is investigated. The photons emerge from a recombination process of charged excitons subsequent to a short, pulsed excitation step. In good approximation the relaxation dynamics can be described by the spontaneous emission of single photons from an ideal two-level system. Accordingly, the photon wave-functions $\zeta_{1,2}\left(t\right)$ can be written as
\begin{equation}
	\zeta_{1,2} \left(t\right) = \frac{1}{\sqrt{\tau_{1,2}}}\cdot \text{H} \left(t\right)\cdot e^{-\frac{t}{2\tau_{1,2}}-\iu 2\pi\nu_{1,2} t},
	\label{wf}
\end{equation}
where $\tau_{1,2}$ denotes the radiative lifetime of the charged exciton state, $\nu_{1,2}$ its instantaneous emission frequency and $\text{H}\left(t\right)$ the Heaviside-function. Moreover, the QDs are subject to spectral diffusion appearing as a jitter of their emission frequencies. Typically, this leads to an inhomogeneous broadening of the spectral line shape following a normal distribution $p_{1,2} \left( \nu_{1,2} \right)$ around the mean frequency $\nu_{\text{c},1,2}$ of the QD with standard deviation $\sigma_{1,2}$ according to
\begin{equation}
	p_{1,2} \left( \nu_{1,2} \right) = \frac{1}{\sigma_{1,2}\sqrt{2\pi}}\cdot\exp{\left[-\frac{1}{2}\left(\frac{\nu_{1,2}-\nu_{\text{c},1,2}}{\sigma_{1,2}}\right)^2\right]}.
	\label{singledist}
\end{equation}
The theoretical description of HOM experiments for the given situation is studied in Ref. \cite{Kambs2018x}. The results therein are used in the present work to assess the remote TPI measurements as shown in \fig{1e}, main article, as well as to predict the tuning curve in \fig{2}, main article. For a better understanding of the underlying equations, the key steps of the derivation shown in Ref.~\cite{Kambs2018x} are briefly summarized in the following.\par
The derivation starts from a well-established formalism describing the HOM experimental situation with the photon fields $\zeta_{1,2}\left(t\right)$ at the two inputs of a BS, respectively \cite{Legero2003x}. Therein, the probability $P\left(t_0,\tau\right)$ with which both input photons leave the BS through distinct output ports and become detected at times $t_0$ and $t_0 + \tau$ is given by
\begin{equation}
	P\left( t_0, \tau \right) = \frac{1}{4}\left|\zeta_1\left(t_0+\tau\right)\zeta_2\left(t_0\right)-\zeta_2\left(t_0+\tau\right)\zeta_1\left(t_0\right)\right|^2.
	\label{pjoint}
\end{equation}
Using the wave-functions (\ref{wf}), the second-order cross-correlation $g^{\left(2\right)}\left(\tau\right)$ can be evaluated to
\begin{align}
	g^{\left(2\right)}\left(\tau\right) &= \int_{-\infty}^{+\infty}{P \left( t_0,\,\tau \right)\,\text{d}t_0} \nonumber \\
	&= \frac{1}{4\left(\tau_1+\tau_2\right)} \times\nonumber \\
	&\left(e^{-\frac{\abs{\tau}}{\tau_1}} + e^{-\frac{\abs{\tau}}{\tau_2}} - 2 \cdot e^{-\frac{\abs{\tau}}{2T}}\cos{2\pi\Delta\nu\tau}\right),
	\label{g2define}
\end{align}
where the instantaneous emission frequency displacement is described by \mbox{$\Delta\nu= \nu_1 - \nu_2$} and \mbox{$1/T = 1/\tau_1 + 1/\tau_2$}. During a long-time measurement $\Delta\nu$ does not stay constant, but is subject to jitter due to the independent spectral diffusion processes of both QDs. For a measurement, which takes much longer than the time both emitters need to explore their frequency ranges (\ref{singledist}), the probability $\rho$ to find a given splitting $\Delta\nu$ is simply given by the cross-correlation of $p_1$ and $p_2$
\begin{align}
	\rho\left(\Delta\nu\right) &= \int_{-\infty}^{+\infty}{p_1\left(\nu\right) p_2\left(\nu + \Delta\nu\right)\,\text{d}\nu}\nonumber \\ &=\frac{1}{\Sigma\sqrt{2\pi}}\cdot\exp{\left[-\frac{1}{2}\left(\frac{\Delta\nu+\delta\nu}{\Sigma}\right)^2\right]}.
\end{align}
Here, $\delta\nu = \nu_{c,1} - \nu_{c,2}$ is the relative displacement of both mean emission frequencies and $\Sigma^2 = \sigma_1^2+\sigma_2^2$ defines the width of $\rho$. Accordingly, the measured cross-correlation $\mathcal{G}^{\left(2\right)}\left(\tau\right)$ in the long-time limit is obtained by a weighted average of \eq{g2define} using $\rho\left(\Delta\nu\right)$ leading to
\begin{align}
	\mathcal{G}^{\left(2\right)}\left(\tau\right)&=\int_{-\infty}^{+\infty}{\rho \left(\Delta\nu\right)g^{\left(2\right)}\left(\tau,\,\Delta\nu\right)\,\text{d}\Delta\nu} \nonumber \\
	&= \frac{1}{4\left(\tau_1+\tau_2\right)} \times\nonumber \\
	&\left(e^{-\frac{\abs{\tau}}{\tau_1}} + e^{-\frac{\abs{\tau}}{\tau_2}} - 2 \cdot e^{-\frac{\abs{\tau}}{2T}}e^{-2\pi^2\Sigma^2\tau^2}\cos{2\pi\delta\nu\tau}\right).
	\label{g2ave}
\end{align}
This is the final result used to describe the central peaks around $\tau= 0$ for all remote TPI correlation measurements shown in the present work.\par
The visibility of HOM-interference is defined by $V = 1 - 2\cdot P$, where $P$ is the overall probability of both photons going separate ways after meeting at the BS. Accordingly, $P$ can be calculated by integrating \eq{g2ave} with respect to the timelag $\tau$. Using the variable $z = \left(2\pi\delta\nu+\iu/2T\right) / \left(2\pi\sqrt{2}\Sigma\right)$, the visibility evaluates to
\begin{equation}
	V = \frac{\text{Re}\left[w\left(z\right)\right]}{\sqrt{2\pi}\Sigma\left(\tau_1+\tau_2\right)}.
	\label{visib}
\end{equation}
where the Faddeeva function $w\left(z\right)$ \cite{abramowitz1964x} is used to express the result. Equation (\ref{visib}) is the definition of a Voigt line shape as a function of the detuning $\delta\nu$, whose width is given by the homogeneous contributions $\tau_{1,2}$ and inhomogeneous contributions $\sigma_{1,2}$. Thus, the visibility of a remote HOM experiment is determined by the joint spectral properties of both emitters. Equation (\ref{visib}) is used in the present work to predict the experimentally achieved visibilities as function of the pump laser detuning for the remote HOM case (see \fig{2}, main text).

\section{Data analyzation}

All correlation histograms are based on time tagging of the raw coincidence events with a \emph{HydraHarp400} with a binning resolution of \SI{1}{\pico\second}. The obtained data is averaged via convolution with the detector response, being $\Delta\tau_\text{SNSPD} \simeq \SI{100}{\pico\second}$ for all data beside the data in Figure 2b-d of the main article where $\Delta\tau_\text{SNSPD} \simeq \SI{40}{\pico\second}$. Furthermore, all data is background corrected.
Each correlation histogram is accompanied with a model which is based on Eq. (\ref{g2ave}). The model is predetermined by the transition lifetimes $\tau_{1,2}$, inhomogeneous broadening, reflected via $\Sigma^2 = \sigma_1^2+\sigma_2^2$, where $\sigma_i = \text{FWHM}_i/(2\sqrt{2\log2})$, and the spectral detuning $\delta\nu_\text{QDs}$ between the two QDs. In order to have an independent access to the actually measured TPI visibility, each correlation measurement has a correlation window being larger than \SI{1}{\micro\second}. This allows to take into account the Poissonian level being the temporally uncorrelated time regime. Comparison between center peak area and the Poissonian level then delivers the TPI visibility (further information on how to access $\VTPI$ can be found in Ref. \cite{Weber2017x}). The error of the measured TPI visibilities is determined via error-propagation, taking the standard errors of center peak and Poissonian level (calculated via $\sqrt N $, where $N$ is the number of coincidence events) as well as the error of background subtraction into account.
\newline
\section{Spectral diffusion}

\begin{figure*}
	\centering
	\includegraphics{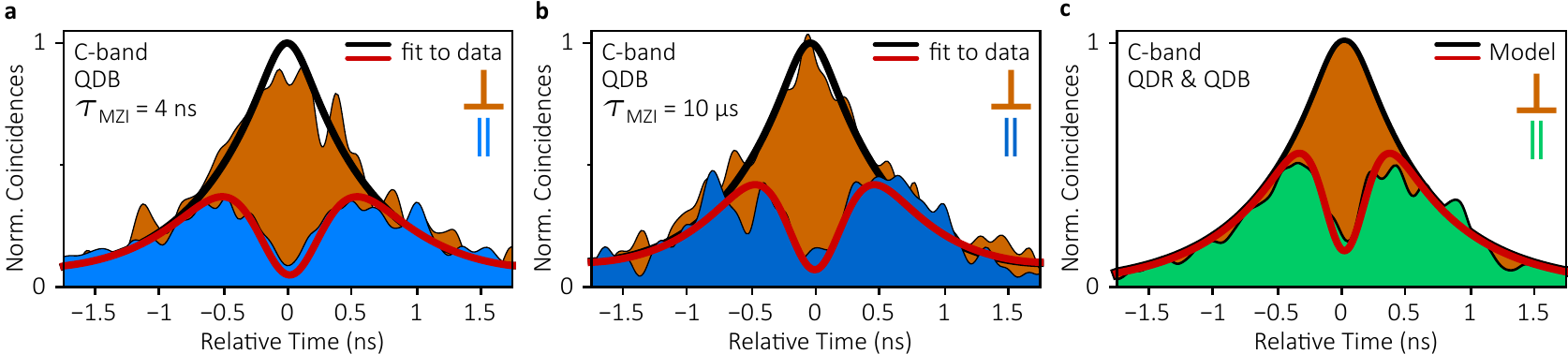}
	\caption{(a,b) Center correlation peak of TPI measurements with consecutively emitted photons of QDB for \SI{4}{\nano\second} and \SI{10}{\micro\second} delay of the MZI normalized to the case with distinguishable photons. (c) Remote TPI measurement of QDR and QDB with parallel and orthogonal polarization setting.}
	\label{fig:figS3}
\end{figure*}

As discussed in course of our theoretical model, the key properties, which drive the indistinguishability of photons from two remote QDs are the integrated emission spectra, including all broadening mechanisms, as well as the lifetime of the addressed state. Spectral diffusion can be quantified by fitting a Voigt profile to the spectral distribution while fixing homogeneous broadening to $\Delta\nu_\text{homog} = \SI{0.27}{\giga\hertz}$, an estimation based on decay times $\tau_\text{dec,QDR}= \SI{580+-10}{\pico\second}$ and $\tau_\text{dec,QDB}= \SI{600+-10}{\pico\second}$ of the two QDs. Actual inhomogeneous broadening can then be extracted, here resulting in $\text{FWHM}_\text{QDR} = \SI{2.0+-0.1}{\giga\hertz}$ and $\text{FWHM}_\text{QDB} = \SI{1.3+-0.1}{\giga\hertz}$. Consequently, time-correlated single-photon counting (TCSPC) and hPL are sufficient to predict the indistinguishability of photons from remote QDs. In contrast, when performing such an experiment with an individual QD it is much more complex to predict the TPI visibility as the spectral diffusion time constant drive the frequency correlation between interfering photons. Figure \ref{fig:figS3}a,b shows TPI measurements with photons subsequently emitted from QDB, however for different setup configuration. The difference between the two measurements comes from a change in time difference $\tau_\text{MZI}$ of the emitted photons. It is realized by inserting \SI{2}{\kilo\meter} of additional fiber path length, hence increasing $\tau_\text{MZI}$ from \SI{4}{\nano\second} to \SI{10}{\micro\second}. As a consequence, the spectral correlation of interfering photons is decreased, i.e. the effective spectral width is approaching the steady state. As the time difference is increased, photon indistinguishability is decreased from $V_\text{QDB}^\text{\SI{4}{\nano\second}}=\SI{61+-3}{\percent}$ to $V_\text{QDB}^\text{\SI{10}{\micro\second}}=\SI{48+-5}{\percent}$. For the case of TPI with remote emitters, this effect is even more prominent as they are spectrally uncorrelated and the visibility reduces to $V_\text{QDR,QDB}^\text{remote}=\SI{29+-3}{\percent}$ (Figure \ref{fig:figS3}c).

\end{document}